\documentclass[reprint,
superscriptaddress,
showpacs,preprintnumbers,
 amsmath,amssymb,
 aps,
 prb,
]{revtex4-1}

\usepackage{multirow}
\usepackage{graphicx}
\usepackage{epstopdf}
\usepackage{placeins}
\usepackage{dcolumn}
\usepackage{bm}
\usepackage{epsfig}
\usepackage{longtable}
\usepackage{cancel}
\usepackage{hyperref}

\usepackage{color}
\definecolor{SBColor}{RGB}{0,0,255} 
\definecolor{SGColor}{RGB}{	255,165,0} 

\begin{document}

\preprint{APS/123-QED}

\title{Bound States of Defects in Superconducting LiFeAs \\ Studied by Scanning Tunneling Spectroscopy}
\author{S. Grothe}
\affiliation{Department of Physics and Astronomy, University of British Columbia, Vancouver BC, Canada V6T 1Z1}
\affiliation{Quantum Matter Institute, University of British Columbia, Vancouver BC, Canada V6T 1Z4}
\author{Shun Chi}
\affiliation{Department of Physics and Astronomy, University of British Columbia, Vancouver BC, Canada V6T 1Z1}
\affiliation{Quantum Matter Institute, University of British Columbia, Vancouver BC, Canada V6T 1Z4}
\author{P. Dosanjh}
\affiliation{Department of Physics and Astronomy, University of British Columbia, Vancouver BC, Canada V6T 1Z1}
\affiliation{Quantum Matter Institute, University of British Columbia, Vancouver BC, Canada V6T 1Z4}
\author{Ruixing Liang}
\affiliation{Department of Physics and Astronomy, University of British Columbia, Vancouver BC, Canada V6T 1Z1}
\affiliation{Quantum Matter Institute, University of British Columbia, Vancouver BC, Canada V6T 1Z4}
\author{W.N. Hardy}
\affiliation{Department of Physics and Astronomy, University of British Columbia, Vancouver BC, Canada V6T 1Z1}
\affiliation{Quantum Matter Institute, University of British Columbia, Vancouver BC, Canada V6T 1Z4}
\author{S. A. Burke}
\affiliation{Department of Physics and Astronomy, University of British Columbia, Vancouver BC, Canada V6T 1Z1}
\affiliation{Department of Chemistry, University of British Columbia, Vancouver BC, Canada V6T 1Z1}
\affiliation{Quantum Matter Institute, University of British Columbia, Vancouver BC, Canada V6T 1Z4}
\author{D. A. Bonn}
\affiliation{Department of Physics and Astronomy, University of British Columbia, Vancouver BC, Canada V6T 1Z1}
\affiliation{Quantum Matter Institute, University of British Columbia, Vancouver BC, Canada V6T 1Z4}
\author{Y. Pennec}
\affiliation{Department of Physics and Astronomy, University of British Columbia, Vancouver BC, Canada V6T 1Z1}
\affiliation{Quantum Matter Institute, University of British Columbia, Vancouver BC, Canada V6T 1Z4}


%


\date{\today}

\begin{abstract}
Defects in LiFeAs are studied by scanning tunneling microscopy and
spectroscopy.
Topographic images of the five predominant defects allow the identification of their position within the lattice.
The most commonly observed defect is associated with an Fe site and does not break the local lattice symmetry, exhibiting a bound state near the edge of the smaller gap in this multi-gap superconductor.  Three other common defects, including one also on an Fe site, are observed to break local lattice symmetry and are pair-breaking indicated by clear in-gap bound states, in addition to states near the smaller gap edge.
STS maps reveal complex, extended real-space bound state patterns, including one with a chiral distribution of the local density of states.
The multiple bound state resonances observed within the gaps and at the inner
gap edge are consistent with theoretical predictions for the s$^{\pm}$ gap symmetry proposed for LiFeAs
and other iron pnictides.
\end{abstract}


\maketitle
\section{Introduction}

Impurity physics plays a key role in superconducting systems, beginning with the remarkable feature that non-magnetic defects do not strongly impact superconductivity in conventional s-wave materials \cite{Anderson1959}. In contrast to such single sign s-wave superconductors, where only magnetic defects cause pair-breaking and in-gap states \cite{Yu1965, *Shiba1968}, both potential and magnetic defects can induce in-gap states in d-wave \cite{Balatsky1995, Yazdani1999PRL, Pan2000Zn, *Hudson2001} and multi-band sign reversal s-wave superconductors (s$^{\pm}$) \cite{Preosti1996, Zhang2009_nonmagnetic, Tsai2009, Kariyado2010}.  Not surprisingly then, a superconductor's sensitivity to defects, plus the energetic and spatial characterization of bound states localized at defect sites have provided clues to the pairing symmetry of novel superconductors, and scanning tunneling microscopy (STM) and spectroscopy (STS)  have proven invaluable tools for such studies
\cite{ Yazdani1997, Yazdani1999PRL, Pan2000Zn, *Hudson2001, Ji2008}. Since a sign change of the order parameter gives rise to sensitivity to defects, it has been suggested that the study of these impurity bound states in the iron arsenides could help close the on-going discussion regarding the gap structure (s$^{++}$ or s$^{\pm}$) \cite{Zhang2009_nonmagnetic, Tsai2009, Kariyado2010}.

The fact that most high temperature superconductors, cuprates as well as pnictides, require chemical substitutions to tune them into their superconducting states adds further importance to understanding the effect of defects in these systems\cite{Bednorz1986, *Kamihara2008}. In the cuprates, this tuning is often achieved through cation substitution on sites away from the CuO$_2$ planes, doping them with holes or electrons, while largely avoiding strong scattering. Direct substitution onto the CuO$_2$ planes is typically pair-breaking, sometimes strongly, sometimes weakly \cite{Alloul2009}.
 For many iron-based superconductors chemical substitution that suppresses extended magnetic order is an essential ingredient in achieving high temperature superconductivity\cite{Wadati2010}.  In BaFe$_2$As$_2$ it has been shown that certain elements such as Co and Ni substituted into the Fe-layer induce superconductivity \cite{Sefat2008, Li2009NiBaFe2As2} while other substituents (Mn) cause strong pair-breaking \cite{Thaler2011Mn}. Thus, the arsenides lack the easy distinction of off-plane substitution to promote superconductivity versus on-plane defects that are pair-breaking. This makes it particularly important to assess individual impurities and their influence on the surrounding electronic states.

STM study of the pnictides has proven difficult due to surface specific effects arising from a lack of natural cleaving planes, or from
structural or electronic reconstruction caused by a polar catastrophe \cite{Hoffman-review-2011}.
Recently, stoichiometric examples within the pnictide and chalcogenide families that exhibit surfaces suitable for STM study such as cleaved LiFeAs crystals \cite{Tapp2008, *Wang2008, *Pitcher2008}, and molecular beam epitaxy grown films of FeSe \cite{Hsu2008} and KFe$_2$Se$_2$ \cite{Li2011} have presented the opportunity to apply STM and STS to well-defined systems.
All three also possess the advantage of being superconducting without chemical substitutions.
In LiFeAs, STM has been used to measure the superconducting gaps of clean defect free areas  \cite{Chi2012, Hanaguri2012}, to study vortices
\cite{Hanaguri2012}, and to determine band structure and gap symmetry through quasiparticle interference induced by defect scattering
\cite{Allan2012, Hess-STM}.
A detailed investigation of the impurities themselves and their localized electronic effects
has yet to be reported, though in-gap states have been observed for iron adatoms in FeSe \cite{Song2011} and possibly iron vacancies in KFe$_2$Se$_2$
\cite{Li2011}. 

In this letter we characterize defects arising from crystal growth in nominally stoichiometric LiFeAs.
We identify their position in the crystal lattice and analyze the spatial and energetic distribution of their bound states.

\section{Experimental Method}
The LiFeAs single crystals were grown by the LiAs self-flux technique.
Li$_{3}$As was pre-synthesized through the reaction of Li (99.9\%) lumps and As (99.9999\%) powder at 773 K for 10 hours. FeAs was pre-synthesized from mixed powders of Fe (99.995\%) and As (99.9999\%) at 973 K for 10 hours. Powders of Li$_{3}$As  and FeAs were mixed in a composition of 1:2 and placed in an alumina crucible, which was sealed under 0.3 atm Ar in a quartz tube. A Mo crucible was used to encapsulate the alumina crucible to prevent Li attack on the quartz tube. All the mixing procedures were done in an Ar atmosphere glovebox. The mixture was heated slowly to 1323 K for 10 hours, then cooled to 1073 K at 4.5 K/hour. Finally, the samples were additionally annealed at 673 K for 12 hours before being removed from the furnace. Single crystals with typical dimensions 2$\times$2$\times$0.2 mm$^3$
were mechanically extracted from the LiAs flux. Lattice parameters $a=(3.777\pm0.004)\text{\AA}$\,  and $c =(6.358\pm0.001)
\,\text{\AA}$\ were determined by x-ray diffraction, and
$T_c^{onset}= 17$ K with a transition width of 1 K was
determined by SQUID magnetometry  with a 1 G magnetic field.

STM/STS measurements were performed in a Createc ultrahigh vacuum low temperature STM. 
The electrochemically etched  tungsten  tip  was  Ar sputtered  and
thermally annealed at the beginning of this experiment. After cleaving the sample in situ at a temperature of 20 K it was immediately transfered to the 4.2 K STM.
The sample is identical to the one used in a previous study \cite{Chi2012},
but it has been recleaved before we obtained the measurements presented here. 
All spectra shown in this report were recorded at a temperature of 2.2 K and 
were acquired by numerical differentiation of the I-V sweep.
All topography scans and dI/dV maps were recorded at 4.2 K.

\begin{figure}[th]
\includegraphics[width=0.49\textwidth]{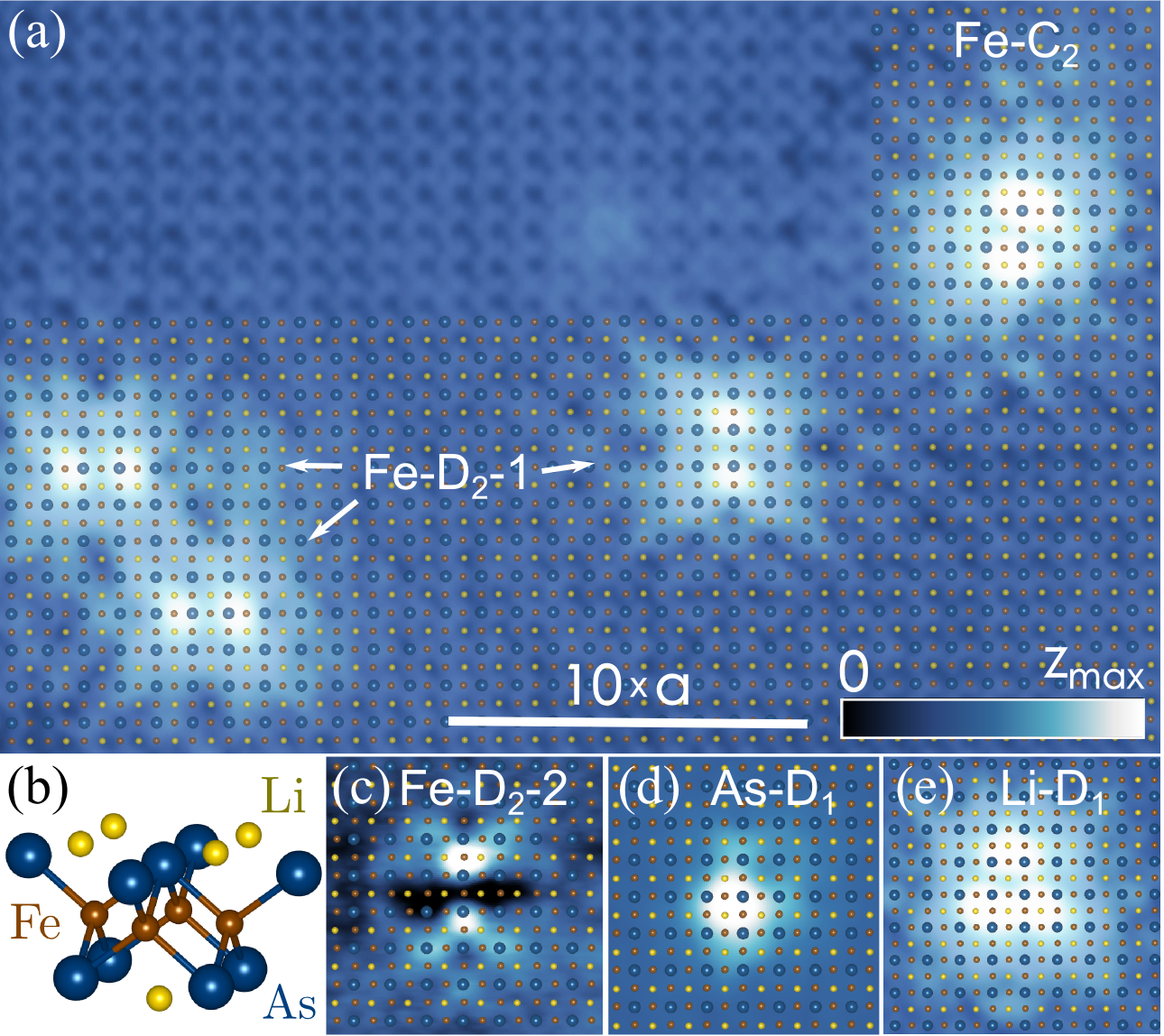}
\caption{\label{fig:registry}
Scanning tunneling topographies of typical defects found at the surface of LiFeAs.
The estimated atomic configuration of the crystal structure with $a=(3.74\pm0.03)$\,\AA\cite{Chi2012} 
is shown on top of
the topographies to locate the defect positions (Li: yellow, As: blue, Fe: red).
(a) 3 different Fe-D$_2$-1 and one Fe-C$_2$ defect. 
(b) Schematic crystal structure in edge-on view of the upper layers of LiFeAs
    with Li at the top as expected from cleaving.
(c), (d) and (e) are topographies of Fe-D$_2$-2, As-D$_1$ and Li-D$_1$ defects.
The tunneling conditions were $I_T=100$ pA for all scans and
$U_B=6$ meV, $z_{max}=120$ pm (a and e), $U_B=-12$ meV, $z_{max}=55$ pm (c),
$U_B=-12$ meV, $z_{max}=120$ pm (d).}
\end{figure}

\section{RESULTS AND DISCUSSION}

Fig. 1 a) and c) to e) show atomic resolution topographic images with common types of defects at the surface of LiFeAs. Defects with similar topography and 
densities have been observed in STM studies of 
LiFeAs grown elsewhere \cite{Allan2012, Hanaguri2012}.
STM scans of LiFeAs also show other types of defects, but we decided to focus on the
most common ones with the largest effect on the electronic structure.
For a clear identification of the Fe, Li and As positions, and thereby assignment of the visible defect sites, a close look at the crystallography and the  defects is needed.

LiFeAs presents a layered crystal structure detailed in Fig. 1 b) \cite{Tapp2008}. At its core is a square lattice of Fe atoms, each nested at the center of a tetrahedron of As. 
After cleaving between the weakly bonded Li layers, the surface consists of a top layer of Li in a square lattice followed by a square lattice of As of the same periodicity but translated by [$1/2, 1/2$]. The third layer is Fe arranged again in a square lattice but of twice the density and rotated $45^\circ$ relative to the Li or As lattice, where neither As nor Li sits directly above or below an Fe site.

The most common defect (labeled Fe-D$_2$-1 in Fig. \ref{fig:registry}a according to labeling principles described below), provides a key clue to the identification of the lattice observed by STM.
Its distinctive feature is the dihedral D$_2$ symmetry exhibiting two bright lobes 
oriented either along [100] or [010]. 
The symmetries discussed in this work are based on the Sch\"onflie\ss~ notation adapted for two dimensions, since STM provides a weighted 2D projection of the 3D crystal most sensitive to
disruptions occurring on the upper rather than the buried planes.

The question is: Which chemical site in the lattice can generate the observed topography?
To form the center of this D$_2$ defect, both the Li and As sites are unlikely as they present D$_4$ symmetry due to the four nearest neighbors within each plane as discussed above.
Iron sites, however, would induce a D$_2$-symmetry in the upper planes due to the tetrahedrally bonded As nearest neighbors 
with two As in the plane above and two in the plane below (Fig. 1b). Therefore, the simplest explanation for this commonly observed defect 
is an iron-site defect, even though other origins such as dimers or interstitial impurities 
cannot be completely excluded. The four strongly polarizable nearest neighbor As \cite{Berciu2009} would be most significantly affected by an iron site defect through charge transfer. 
But STM is much more sensitive to changes on the upper two As above the Fe layer, 
resulting in a prominent charge density contrast along the direction of these two As atoms.
Likely this defect arises from Fe vacancies or Li substitution of iron sites, as the crystal was grown from a Li-rich flux. 
In the following we refer to this defect as Fe-D$_2$-1 based on the apparent center position and symmetry.

The spatial assignment of the Fe-D$_2$-1 defect, centered on an iron site and extended along the line of the two  As nearest neighbors lying above the Fe layer,
allows the registry of the three atomic sublattices (Fe, As and Li) to the measured STM corrugation.
 In Fig. 1a) the STM topography is overlaid with a model of the (001) cleaved crystal structure of LiFeAs
(Li: yellow, As: blue, and Fe: orange), where the Fe site sits at the center of the Fe-D$_2$-1 defect and the two topmost As are located at the lobes.
This model works remarkably well across the full scan with no phase shift between the applied grid and the measured atomic periodicity matching the Fe periodicity.

Previously, atomic resolution STM images of LiFeAs have produced the periodicity of either Li or As \cite{Hess-STM, Chi2012, Allan2012, Hanaguri2012}.
Here, the periodicity of the Fe lattice is observed at low bias voltages ($|U_B|\lesssim 20$ meV) and with a particular tip wavefunction, which also causes to a stronger weight of the smaller energy gap in STS. 
A typical STS spectrum taken at the center of a defect-free area is shown in Fig. 2a) in solid black. As reported previously \cite{Chi2012, Hanaguri2012, Allan2012}, two nodeless gaps are clearly resolved with half width peak-to-peak of $\Delta^{pp}_1=6$ meV and half width shoulder to shoulder $\Delta^{pp}_2=3$ meV, but the spectral weight contribution of the small gap $\Delta_2$ is about twice as large as previously found \cite{Chi2012}.  
The STM quasiparticle interference study by Allan \textit{et al.} identified the small energy gap
with a size of about 2 to 3 meV and a negative dispersion as being associated with the outer hole pocket \cite{Allan2012},
which also corresponds to the in-plane Fe-d$_{xy}$-orbital \cite{Borisenko2012, Ferber2012, Hajiri2012}. 
Tunneling into the electron
pockets which have d$_{xy}$, d$_{xz}$ and d$_{yz}$ character and which contain the other two  gaps \cite{Borisenko2012, Ferber2012, Hajiri2012}, cannot be  completely excluded but is expected to be strongly suppressed because of the larger in-plane momentum $|\vec k_{||}|$ \cite{Tersoff1998}.
Thus we conclude that the sensitivity to
the iron corrugation observed here is combined with an enhanced tunneling into the iron in-plane d$_{xy}$ orbital.
\footnote{The energetic positions of bound-state resonances are not expected to be
influenced by the tip state, which we have confirmed by comparing the data shown here for 
the Fe-D$_2$-1 with 4 K spectra on this defect from our previous data \cite{Chi2012} showing the lower weight of $\Delta_2$.}

\begin{figure}[tbh]
\includegraphics[width=0.5\textwidth]{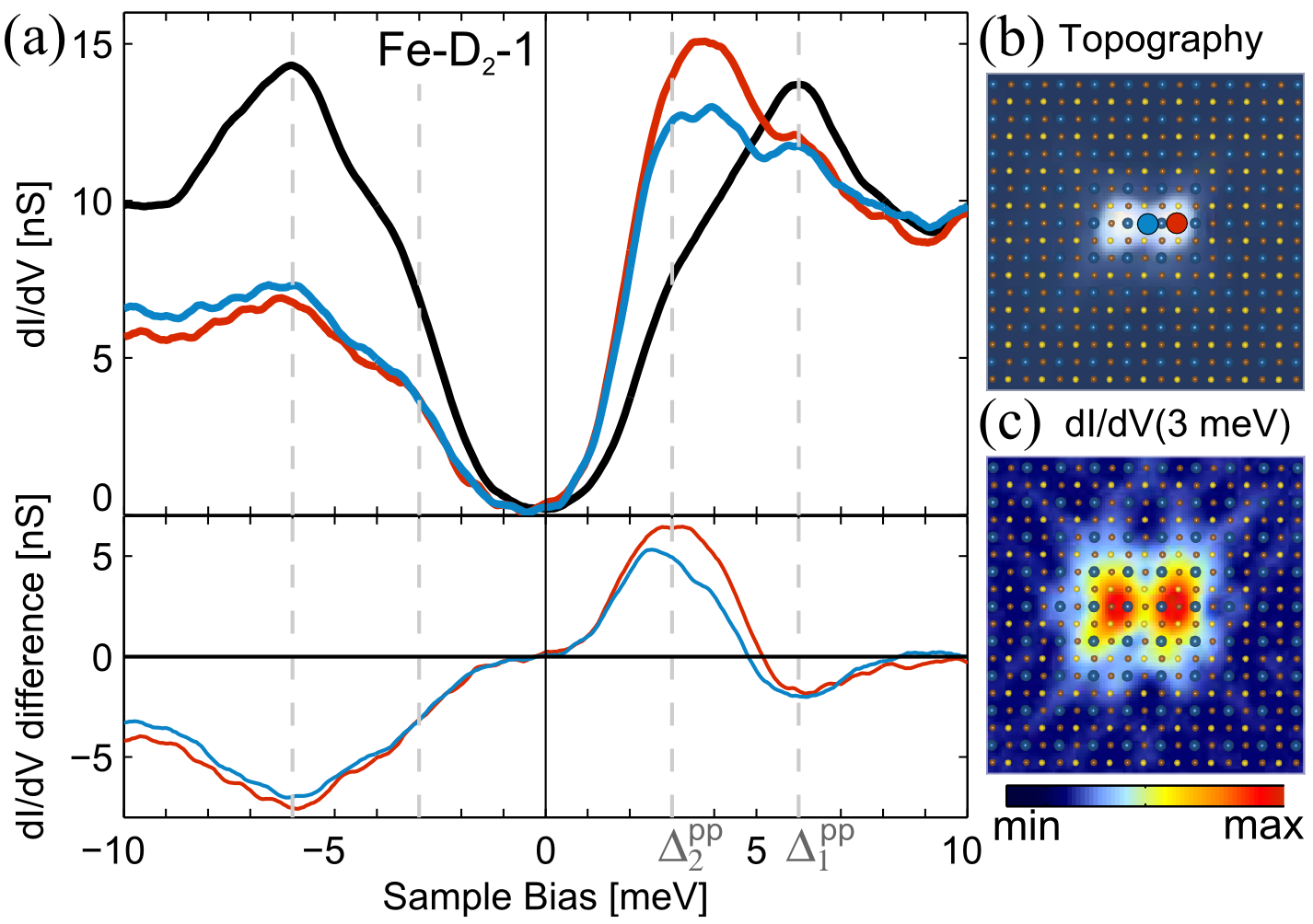}
\caption{\label{fig:FeD2}
 a) The top panel shows $dI/dV$ spectra taken at a Fe-D$_2$-1 defect (red and blue lines) and
 the average over 20 spectra measured about 2 nm away from the defect (black line).
 The thin red and blue lines in the lower panel correspond to the difference between the spectra taken at the defect
 and the average from the defect-free position.
 A bound state is visible at approximately 3 meV.
 Spectra were taken at the positions marked by single red and blue dots in the topographic image b).
 c) $dI/dV$ map at an energy of $3\pm 1$ meV, i.e. averaged from 2 meV to 4 meV.
 The topography and the $dI/dV$ maps are overlayed with the top-view crystal structure.
 Spectra shown in a) were taken at a temperature of 2.2 K.
 Topography and $dI/dV$ map shown in b) and c) were recorded at 4.2 K with a constant height defined by
 $U_{B,0}=25$ meV, $I_0= 260$ pA.
 Topography and $dI/dV$ maps have edge lengths of $\approx$  3.4 nm.
}
\end{figure}

Turning to the effect of the Fe-D$_2$-1 on the local density of states (LDOS) in the vicinity of the defect, the energetic and spatial distributions of bound states are revealed.
Fig. 2a) upper panel shows two raw spectra taken at 2.2 K at the center and on the lobes of the Fe-D$_2$-1 defect as well as a reference spectrum of pristine LiFeAs, taken from the same region. The lower panel presents
these two spectra normalized by subtracting the reference to enhance subtle features \footnote{
We note that 
this normalization procedure may slightly shift energetic features especially if there is an overall redistribution of spectral weight, a gap suppression, or multiple peaks that are within our 200 $\mu$V resolution. Since these factors likely exceed the small statistical uncertainty of locating the peak position, we report only approximate values for bound state energies.}.
The first effect of the defect is a suppression of spectral weight at the large gap coherence peaks. More interestingly, both spectra reveal a resonance at 3 meV coinciding with the edge of the small gap $\Delta_2^{pp}$
and slightly more pronounced at the lobe than at the defect center.
$\Delta_2$ is on a Fermi surface segment dominated by the iron in-plane d$_{xy}$ band, expected to be
influenced by defects in the iron plane. A bias symmetric counterpart at -3 meV is not visible within our resolution.
The 4.2 K spatial distribution of the 3 meV bound state resonance, shown in Fig. 2 c), follows the D$_2$ pattern measured in topography.
To enhance spatial features, the maps were averaged over the 2 meV width
of the bound state, i.e. from 2 meV to 4 meV for the 3 meV map.

We now address four other recurrent defects of LiFeAs which we register based on our identification of the lattice as shown in Fig \ref{fig:registry}a), and c-e).
A second Fe-centered defect also has a dihedral D$_2$ symmetry
and is referred to as Fe-D$_2$-2 (Fig. 1c).
Fe-D$_2$-2, which we will not discuss in detail, exhibits similar electronic properties as Fe-D$_2$-1 but with a weaker 3 meV bound state resonance.
The others are labeled Fe-C$_2$, Li-D$_1$ and As-D$_1$ based on the apparent defect site registry in the xy-plane and their two dimensional group symmetry.  
We note that the center position of the Li-D$_1$ and As-D$_1$ defects is somewhat ambiguous due to their large and complex spatial extents. Here the notation gives the regular lattice position closest to the apparent defect center.
We observed the two possible chiralities and two orientations of Fe-C$_2$ as well as all four possible orientations of Li-D$_1$ and As-D$_1$. We note that except for the two Fe-D$_2$ defects, the other three break the local symmetry of the crystal lattice. This symmetry breaking can be caused by particular orbital orientations close to the defect \cite{Amara2007}, by interstitial impurity positions or by the occurrence of two or more impurities close to each other.
A summary of the defect characteristics is presented in Table-\ref{tab:summary}.

\begin{figure}[tb]
\includegraphics[width=0.49\textwidth]{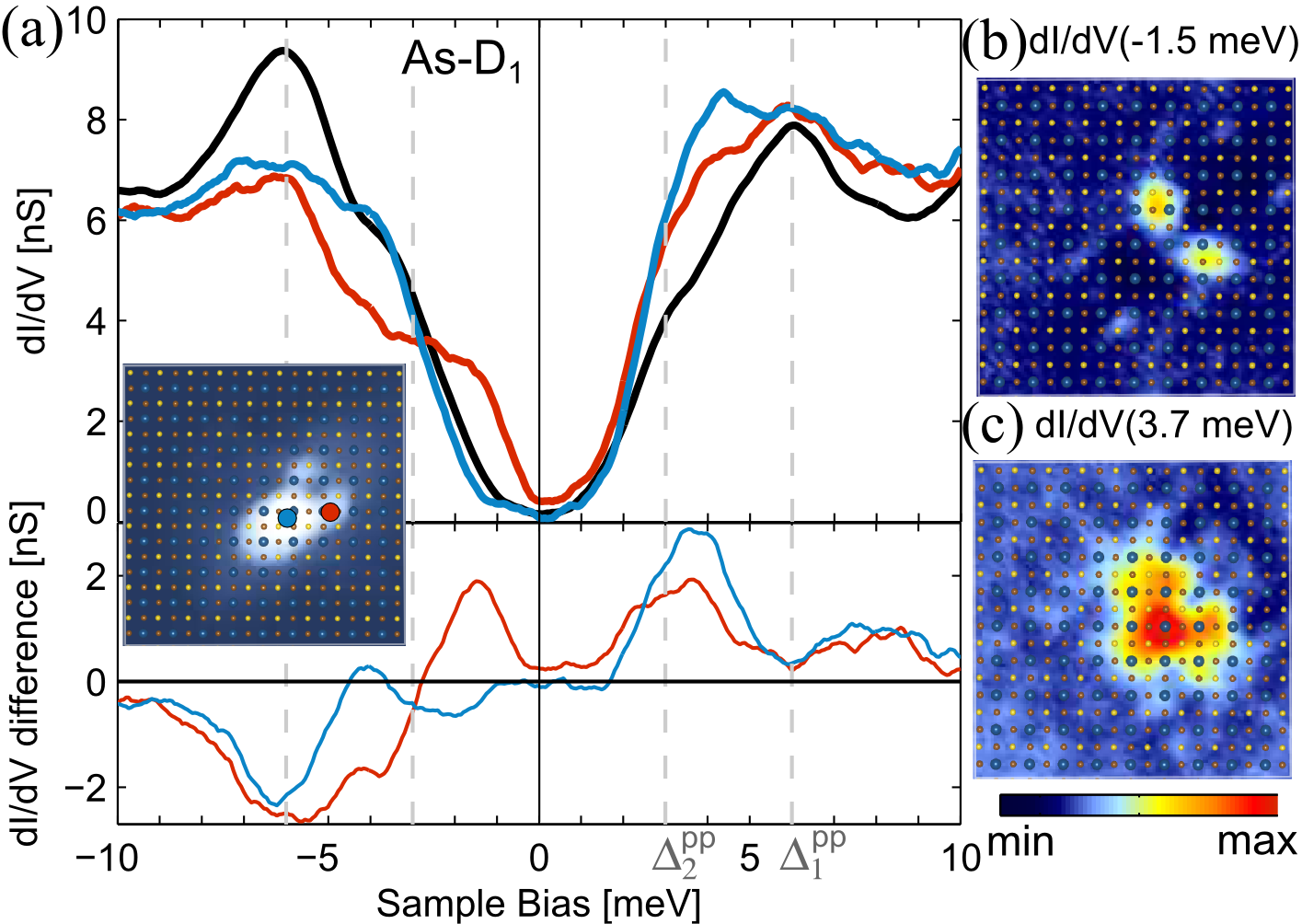}
\caption{\label{fig:AsD1}
 Same as Fig. \ref{fig:FeD2} but for the As-D$_1$ defect.
 Bound states are visible at approximately -1.5 meV and at + 3.7 meV.
 The topography which indicates the location of the spectra is shown as an inset of a).
 b) and c) $dI/dV$ maps at energies of $-1.5$ and $3.7\pm 1$ meV, respectively.}
\end{figure}

Fig 3a) shows spectra taken at the center and the arm of an As-D$_1$ defect. Similar to Fe-D$_2$-1, As-D$_1$ causes a strong suppression of the coherence peaks of $\Delta_1$. However, the spectra not only reveal a resonance close to the positive edge of $\Delta_2$ but also indicate a peak below -$\Delta_2$ at about $-3.7$ meV.
Additionally, another resonance appears inside the gap at about -1.5 meV. The -1.5 meV resonance is present only in the arm of the defect but not at its center. This is made more clear by the spectral maps acquired at -1.5 meV and 3.7 meV presented on Fig. 3b) and 3c).

\begin{figure}[tb]
\includegraphics[width=0.49\textwidth]{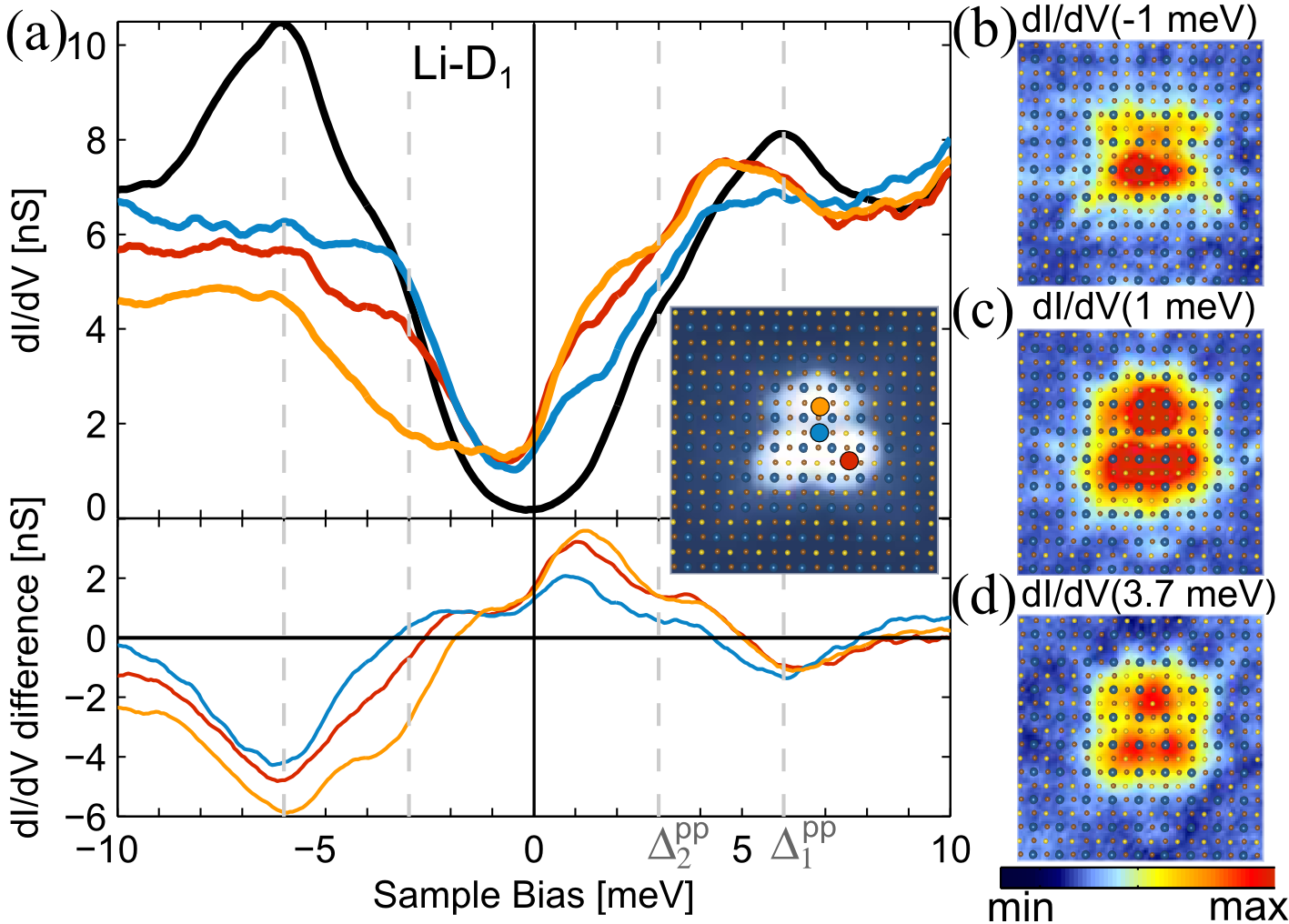}
\caption{\label{fig:LiD1}
 Same as Fig. \ref{fig:FeD2} but for the Li-D$_1$ defect.
 a) A bound state is visible at approximately + 1.2 meV.
  The topography which indicates the location of the spectra is shown as an inset of a).
 b), c) and d) show $dI/dV$ maps at energies of -1, 1 and 3.7 $\pm$ 1 meV, respectively.
}
\end{figure}

Spectra taken on the Li-D$_1$ defect are shown in Fig. 4a). Again, the coherence peaks of the large gap are reduced but a resonance near $\Delta_2$ is just weakly suggested.
The three spectra in Fig. 4a) are qualitatively similar, showing a clear in-gap resonance at about 1.2 meV. Spectral maps at 1 meV, at the bias symmetric energy
-1 meV and at the 3.7 meV peak just above $\Delta_2$ are shown in Fig. 4b), c) and d), revealing that the near 1 meV state is the most extended.
At -1 meV, the LDOS is predominantly localized on two As sites near the Li center.

\begin{figure}[b]
\includegraphics[width=0.49\textwidth]{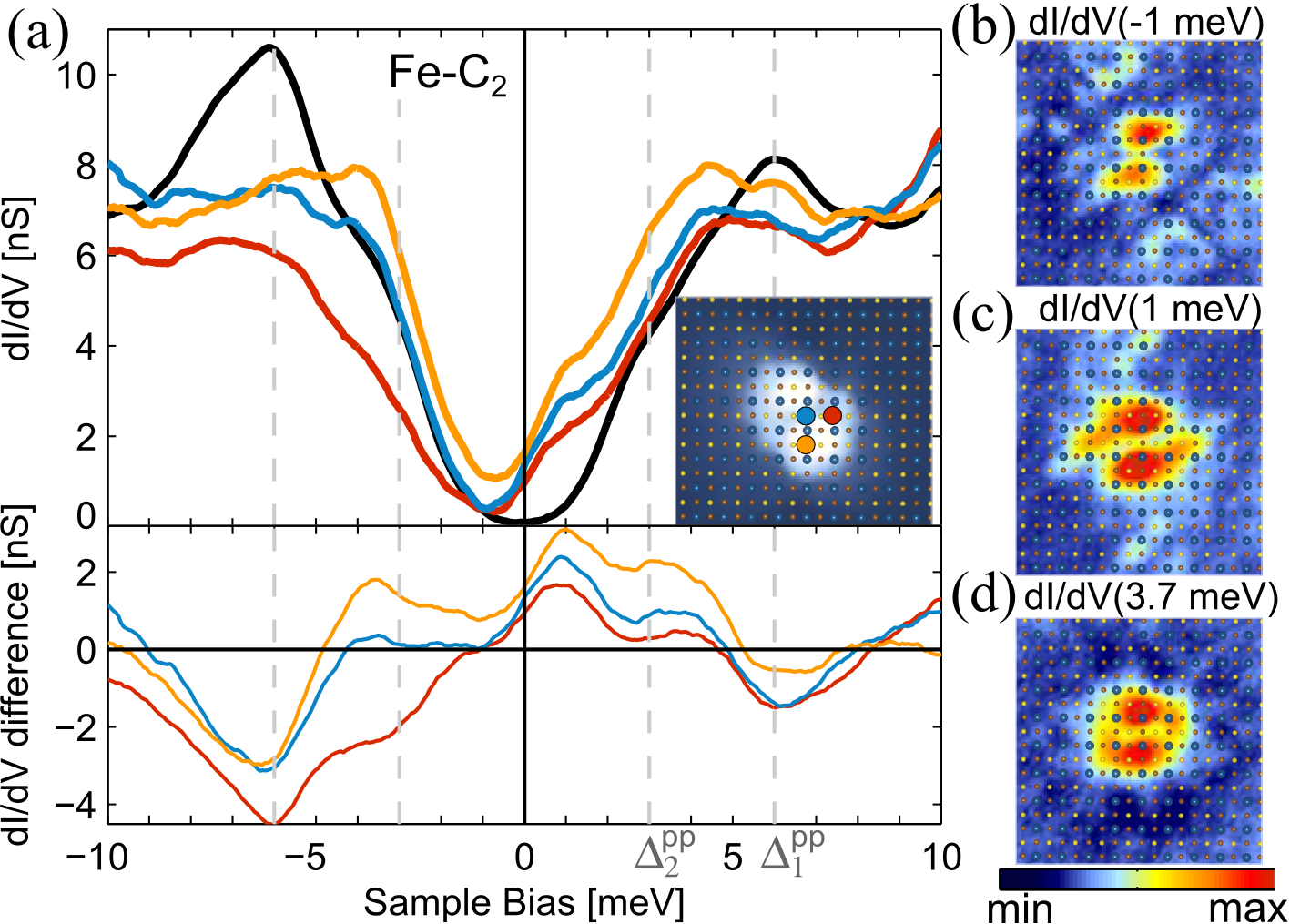}
\caption{\label{fig:FeC2}
 Same as Fig. \ref{fig:FeD2} but for the Fe-C$_2$ defect.
 a) Bound states are visible at approximately + 1.0 meV and at about $\pm$ 3.5 meV.
  The topography which indicates the location of the spectra is shown as an inset of a).
 b), c) and d) show $dI/dV$ maps at energies of -1, 1 and 3.7 $\pm 1$ meV, respectively.
}
\end{figure}

Fig. 5a) shows spectra recorded on three sites of a Fe-C$_2$ defect. Once more, the defect reduces the amplitude of the coherence peaks of $\Delta_1$. Resonances at 1 meV and approximately $\pm$ 3.5 meV are observed.
The spatial $dI/dV$ mapping of these defects differs from the constant current map topography.
The $dI/dV$ map at the -3.5 meV resonance (not shown) is qualitatively similar to the -1 meV map.
Interestingly, the $-1$ meV and 3.7 meV maps show one chirality while the opposite chirality is observed at $+1$ meV.

\begin{table}[t]
\caption{\label{tab:summary} Summary of  the five defect types. 
The approximate density is the number per LiFeAs formula unit and has
been obtained for one sample cleaved two times. Other samples have shown the same types of defects, with similar relative densities, but sample-to-sample differences of absolute densities.
$E_b$ is the bound state energy relative to $E_F$. 
(*) indicates those defects where the spatial assignment
is ambiguous.}
\begin{ruledtabular}
\begin{tabular}{cccccc}
 Defect   &Symmetry&Density    & $E_b$ & \multirow{2}{*}{$\displaystyle \frac{\left|E_b\right|}{\Delta^{pp}_1}$}&\multirow{2}{*}{$\displaystyle \frac{\left| E_b\right|}{\Delta^{pp}_2}$}\\
       &         &[$10^{-3}$]& [meV] &                                   &                                    \\ \hline
 Fe-D$_2$-1&D$_2$&$\gtrsim 1.2$&$\approx$  3.0&0.5&1\\
 Fe-D$_2$-2&D$_2$&$\lesssim 0.2$&$\approx$  3.0&0.5&1\\
 As-D$_1$(*) &D$_1$&$\lesssim 0.5$&$\approx$   3.7&0.62&1.23\\
                                            &&&$\approx$-1.5&0.25&0.5\\
 Li-D$_1$ (*)&D$_1$&$\lesssim 0.1$&$\approx$    3.7&0.62&1.23\\
                                          &&& $\approx$ 1.2&0.2&0.4\\
 Fe-C$_2$&C$_2$&$\lesssim 0.1$&$\approx\pm 3.5$&0.58&1.17\\
          &&&$\approx$    1.0&0.17&0.33\\
\end{tabular}
\end{ruledtabular}
\end{table}

Summarizing the properties of all defects,
some striking features of the $dI/dV$ maps presented here should be highlighted.
Bound state resonances exhibit large spatial extents, most distinct within an area of $\sim4\times4$ unit cells.
Three of the five defect types observed in LiFeAs break the lattice symmetry and the bound state patterns are not commensurate with the crystal lattice.
The strong dependence on the bias voltage reveals different spatial localization
of different resonance energies.
Consequently, bound states in LiFeAs result in more complex patterns than those observed in the cuprates where the  LDOS of particle and hole components shows complementary patterns overlaying the CuO lattice sites \citep{Pan2000Zn,Hudson2001}.
In particular, the chiral C$_2$ pattern of the Fe-C$_2$ defect is rarely observed.  
A recent similar observation in NbSe$_2$ is attributed to charge density waves \cite{Ishioka2011}, and
it has been recently proposed that impurities in pnictides may yield chiral C$_2$ LDOS patterns in the presence of spin density order \cite{Huang2011} or strong orbital fluctuations \cite{Inoue2012}.
If the extraordinary chiral pattern of the Fe-C$_2$ defect is indeed connected to magnetic, charge or orbital order locally frozen by the defect, the present observations might provide further insight regarding the pairing glue in LiFeAs.

We turn to the analysis of bound state energies relative to the gap sizes, which can help to
identify the gap structure.
Both of the two lattice symmetry preserving Fe-D$_2$ defects stand out as having only one bound state appearing near the edge of the small gap. As-D$_2$, Li-D$_1$ and Fe-C$_2$ also cause resonances at or near the small gap edge $\Delta_2$,
but lower temperature measurements are needed to clearly resolve these features.
As-D$_2$, Li-D$_1$ and Fe-C$_2$, which break the local lattice symmetry, induce pair-breaking indicated by clear in-gap bound states pronounced at either positive or negative bias.
For most resonances only one peak is visible and
the expected bias symmetric counterpart \citep{Zhang2009_nonmagnetic,Tsai2009,Kariyado2010}
is either not observed or only weakly present, revealing a significant particle-hole asymmetry of the bound states as observed by STS.
Due to our resolution and the uncertainty inherent in finding the bound state peaks on a background that is also influenced by the presence of the defect, we cannot definitively ascribe these states to the small gap edge, and cannot exclude that they are in-gap states of the larger gap.  However, the asymmetric multi-peak energetic structure we observe and the observation of states near the small gap are consistent with recent theoretical predictions for s$^\pm$ symmetries \cite{Zhang2009_nonmagnetic, Kariyado2010}. 
While we cannot exclude the possibility that the defects in the crystal exhibit magnetic properties, potentially inducing in-gap bound states for either  s$^\pm$ and s$^{++}$ \cite{Zhang2009_nonmagnetic, Tsai2009}, such states at the gap edge have not previously been observed in conventional s-wave superconductors \cite{Balatsky2006review,Yazdani1997,Ji2008}.
Controlled introduction of impurities and probing of the bound state resonances should provide further evidence for the pairing symmetry in the iron pnictides.

Regardless of pairing symmetry and interaction, we gain information regarding the resilience of the superconductive phase of LiFeAs to chemical substitution within the iron layer, reminiscent of the observation of pair-breaking and non-pair breaking Fe-site defects in BaFe$_2$As$_2$ \cite{Sefat2008, Li2009NiBaFe2As2, Thaler2011Mn}. Even though all common LiFeAs defects show bound states within the large gap $\Delta_1$, the two Fe-D$_2$ defects have no states within $\Delta_2$, raising the question of how defects affect charge carriers in different bands. In contrast, Fe-C$_2$, with its bound state below both gaps, clearly corresponds to a pair-breaking substituent also likely located within the iron layer.

\section{Conclusion}

We classify and register the common defects appearing at the surface of LiFeAs.
Our registry can be used as a guideline for the
identification of deliberately introduced external impurities.
The bound states show complex spatial patterns that vary strongly with energy and
do not reflect the lattice periodicity.
Further, the energetic multi-peak bound state structures with resonances at the small superconducting gap
are in accordance with calculations of non-magnetic defects in s$^{\pm}$ gap structures \cite{Kariyado2010},
although the possibility of a magnetic nature of our defects does not allow a definitive attribution of the gap structure.
Finally, the observed variety of different defects suggests many possibilities for controlled doping
and thereby for tailoring the material properties.

\acknowledgements
The authors would like to thank Ilya Elfimov, George Sawatzky and Giorgio Levy for several helpful conversations.
This work was supported by the Canadian Institute for Advanced Research, the Canada Foundation
for Innovation, and the Natural Sciences and Engineering Research Council of Canada.

\bibliographystyle{apsrev4-1}
\bibliography{Defect-paper-Sept-04}

\begin{thebibliography}{43}%
\makeatletter
\providecommand \@ifxundefined [1]{%
 \@ifx{#1\undefined}
}%
\providecommand \@ifnum [1]{%
 \ifnum #1\expandafter \@firstoftwo
 \else \expandafter \@secondoftwo
 \fi
}%
\providecommand \@ifx [1]{%
 \ifx #1\expandafter \@firstoftwo
 \else \expandafter \@secondoftwo
 \fi
}%
\providecommand \natexlab [1]{#1}%
\providecommand \enquote  [1]{``#1''}%
\providecommand \bibnamefont  [1]{#1}%
\providecommand \bibfnamefont [1]{#1}%
\providecommand \citenamefont [1]{#1}%
\providecommand \href@noop [0]{\@secondoftwo}%
\providecommand \href [0]{\begingroup \@sanitize@url \@href}%
\providecommand \@href[1]{\@@startlink{#1}\@@href}%
\providecommand \@@href[1]{\endgroup#1\@@endlink}%
\providecommand \@sanitize@url [0]{\catcode `\\12\catcode `\$12\catcode
  `\&12\catcode `\#12\catcode `\^12\catcode `\_12\catcode `\%12\relax}%
\providecommand \@@startlink[1]{}%
\providecommand \@@endlink[0]{}%
\providecommand \url  [0]{\begingroup\@sanitize@url \@url }%
\providecommand \@url [1]{\endgroup\@href {#1}{\urlprefix }}%
\providecommand \urlprefix  [0]{URL }%
\providecommand \Eprint [0]{\href }%
\providecommand \doibase [0]{http://dx.doi.org/}%
\providecommand \selectlanguage [0]{\@gobble}%
\providecommand \bibinfo  [0]{\@secondoftwo}%
\providecommand \bibfield  [0]{\@secondoftwo}%
\providecommand \translation [1]{[#1]}%
\providecommand \BibitemOpen [0]{}%
\providecommand \bibitemStop [0]{}%
\providecommand \bibitemNoStop [0]{.\EOS\space}%
\providecommand \EOS [0]{\spacefactor3000\relax}%
\providecommand \BibitemShut  [1]{\csname bibitem#1\endcsname}%
\let\auto@bib@innerbib\@empty
\bibitem [{\citenamefont {Anderson}(1959)}]{Anderson1959}%
  \BibitemOpen
  \bibfield  {author} {\bibinfo {author} {\bibfnamefont {P.~W.}\ \bibnamefont
  {Anderson}},\ }\href {\doibase 10.1016/0022-3697(59)90036-8} {\bibfield
  {journal} {\bibinfo  {journal} {J. Phys. Chem. Solids}\ }\textbf {\bibinfo
  {volume} {11}},\ \bibinfo {pages} {26} (\bibinfo {year} {1959})}\BibitemShut
  {NoStop}%
\bibitem [{\citenamefont {{L. Yu}}(1965)}]{Yu1965}%
  \BibitemOpen
  \bibfield  {author} {\bibinfo {author} {\bibnamefont {{L. Yu}}},\ }\href
  {http://wulixb.iphy.ac.cn/EN/abstract/abstract851.shtml} {\bibfield
  {journal} {\bibinfo  {journal} {Acta Phys. Sin.}\ }\textbf {\bibinfo {volume}
  {21}} (\bibinfo {year} {1965})}\BibitemShut {NoStop}%
\bibitem [{\citenamefont {{H. Shiba}}(1968)}]{Shiba1968}%
  \BibitemOpen
  \bibfield  {author} {\bibinfo {author} {\bibnamefont {{H. Shiba}}},\ }\href
  {\doibase 10.1143/PTP.40.435} {\bibfield  {journal} {\bibinfo  {journal}
  {Prog. Theor. Phys}\ }\textbf {\bibinfo {volume} {40}},\ \bibinfo {pages}
  {435} (\bibinfo {year} {1968})}\BibitemShut {NoStop}%
\bibitem [{\citenamefont {{A. V. Balatsky, M. I. Salkola and A.
  Rosengren}}(1995)}]{Balatsky1995}%
  \BibitemOpen
  \bibfield  {author} {\bibinfo {author} {\bibnamefont {{A. V. Balatsky, M. I.
  Salkola and A. Rosengren}}},\ }\href {\doibase 10.1103/PhysRevB.51.15547}
  {\bibfield  {journal} {\bibinfo  {journal} {Phys. Rev. B}\ }\textbf {\bibinfo
  {volume} {51}},\ \bibinfo {pages} {15547} (\bibinfo {year}
  {1995})}\BibitemShut {NoStop}%
\bibitem [{\citenamefont {{A. Yazdani, C. M. Howald, C. P. Lutz, A. Kapitulnik
  and D. M. Eigler}}(1999)}]{Yazdani1999PRL}%
  \BibitemOpen
  \bibfield  {author} {\bibinfo {author} {\bibnamefont {{A. Yazdani, C. M.
  Howald, C. P. Lutz, A. Kapitulnik and D. M. Eigler}}},\ }\href {\doibase
  10.1103/PhysRevLett.83.176} {\bibfield  {journal} {\bibinfo  {journal} {Phys.
  Rev. Lett.}\ }\textbf {\bibinfo {volume} {83}},\ \bibinfo {pages} {176}
  (\bibinfo {year} {1999})}\BibitemShut {NoStop}%
\bibitem [{\citenamefont {{S. H. Pan, E. W. Hudson, K. M. Lang, H. Eisaki, S.
  Uchida and J. C. Davis}}(2000)}]{Pan2000Zn}%
  \BibitemOpen
  \bibfield  {author} {\bibinfo {author} {\bibnamefont {{S. H. Pan, E. W.
  Hudson, K. M. Lang, H. Eisaki, S. Uchida and J. C. Davis}}},\ }\href
  {\doibase 10.1038/35001534} {\bibfield  {journal} {\bibinfo  {journal}
  {Nature}\ }\textbf {\bibinfo {volume} {403}},\ \bibinfo {pages} {746}
  (\bibinfo {year} {2000})}\BibitemShut {NoStop}%
\bibitem [{\citenamefont {{E. W. Hudson, K. M. Lang, V. Madhavan, S. H. Pan, H.
  Eisaki, S. Uchida and J. C. Davis}}(2001)}]{Hudson2001}%
  \BibitemOpen
  \bibfield  {author} {\bibinfo {author} {\bibnamefont {{E. W. Hudson, K. M.
  Lang, V. Madhavan, S. H. Pan, H. Eisaki, S. Uchida and J. C. Davis}}},\
  }\href {\doibase 10.1038/35082019} {\bibfield  {journal} {\bibinfo  {journal}
  {Nature}\ }\textbf {\bibinfo {volume} {411}},\ \bibinfo {pages} {920}
  (\bibinfo {year} {2001})}\BibitemShut {NoStop}%
\bibitem [{\citenamefont {{G. Preosti and P. Muzikar}}(1996)}]{Preosti1996}%
  \BibitemOpen
  \bibfield  {author} {\bibinfo {author} {\bibnamefont {{G. Preosti and P.
  Muzikar}}},\ }\href {\doibase 10.1103/PhysRevB.54.3489} {\bibfield  {journal}
  {\bibinfo  {journal} {Phys. Rev. B}\ }\textbf {\bibinfo {volume} {54}},\
  \bibinfo {pages} {3489} (\bibinfo {year} {1996})}\BibitemShut {NoStop}%
\bibitem [{\citenamefont {Zhang}(2009)}]{Zhang2009_nonmagnetic}%
  \BibitemOpen
  \bibfield  {author} {\bibinfo {author} {\bibfnamefont {D.}~\bibnamefont
  {Zhang}},\ }\href {\doibase 10.1103/PhysRevLett.103.186402} {\bibfield
  {journal} {\bibinfo  {journal} {Phys. Rev. Lett.}\ }\textbf {\bibinfo
  {volume} {103}},\ \bibinfo {pages} {186402} (\bibinfo {year}
  {2009})}\BibitemShut {NoStop}%
\bibitem [{\citenamefont {{W.-F. Tsai, Y.-Y. Zhang, C. Fang, and J.
  Hu}}(2009)}]{Tsai2009}%
  \BibitemOpen
  \bibfield  {author} {\bibinfo {author} {\bibnamefont {{W.-F. Tsai, Y.-Y.
  Zhang, C. Fang, and J. Hu}}},\ }\href {\doibase 10.1103/PhysRevB.80.064513}
  {\bibfield  {journal} {\bibinfo  {journal} {Phys. Rev. B}\ }\textbf {\bibinfo
  {volume} {80}},\ \bibinfo {pages} {064513} (\bibinfo {year}
  {2009})}\BibitemShut {NoStop}%
\bibitem [{\citenamefont {Kariyado}\ and\ \citenamefont
  {Ogata}(2010)}]{Kariyado2010}%
  \BibitemOpen
  \bibfield  {author} {\bibinfo {author} {\bibfnamefont {T.}~\bibnamefont
  {Kariyado}}\ and\ \bibinfo {author} {\bibfnamefont {M.}~\bibnamefont
  {Ogata}},\ }\href {\doibase 10.1143/JPSJ.79.083704} {\bibfield  {journal}
  {\bibinfo  {journal} {J. Phys. Soc. Jpn.}\ }\textbf {\bibinfo {volume}
  {79}},\ \bibinfo {pages} {083704} (\bibinfo {year} {2010})}\BibitemShut
  {NoStop}%
\bibitem [{\citenamefont {{A. Yazdani, B. A. Jones, C. P. Lutz, M. F. Crommie,
  D. M. Eigler}}(1997)}]{Yazdani1997}%
  \BibitemOpen
  \bibfield  {author} {\bibinfo {author} {\bibnamefont {{A. Yazdani, B. A.
  Jones, C. P. Lutz, M. F. Crommie, D. M. Eigler}}},\ }\href {\doibase
  10.1126/science.275.5307.1767} {\bibfield  {journal} {\bibinfo  {journal}
  {Science}\ }\textbf {\bibinfo {volume} {275}},\ \bibinfo {pages} {1767}
  (\bibinfo {year} {1997})}\BibitemShut {NoStop}%
\bibitem [{\citenamefont {{S.-H. Ji, T. Zhang, Y.-S. Fu, X. Chen, X.-C. Ma, J.
  Li, W.-H. Duan, J.-F. Jia, and Q.-K. Xue}}(2008)}]{Ji2008}%
  \BibitemOpen
  \bibfield  {author} {\bibinfo {author} {\bibnamefont {{S.-H. Ji, T. Zhang,
  Y.-S. Fu, X. Chen, X.-C. Ma, J. Li, W.-H. Duan, J.-F. Jia, and Q.-K. Xue}}},\
  }\href {\doibase 10.1103/PhysRevLett.100.226801} {\bibfield  {journal}
  {\bibinfo  {journal} {Phys. Rev. Lett.}\ }\textbf {\bibinfo {volume} {100}},\
  \bibinfo {pages} {226801} (\bibinfo {year} {2008})}\BibitemShut {NoStop}%
\bibitem [{\citenamefont {{J. G. Bednorz and K. A.
  M{\"u}ller}}(1986)}]{Bednorz1986}%
  \BibitemOpen
  \bibfield  {author} {\bibinfo {author} {\bibnamefont {{J. G. Bednorz and K.
  A. M{\"u}ller}}},\ }\href {\doibase 10.1007/BF01303701} {\bibfield  {journal}
  {\bibinfo  {journal} {Z. Phys. B}\ }\textbf {\bibinfo {volume} {64}},\
  \bibinfo {pages} {189} (\bibinfo {year} {1986})}\BibitemShut {NoStop}%
\bibitem [{\citenamefont {{Y. Kamihara, T. Watanabe, M. Hirano, and H.
  Hosono}}(2008)}]{Kamihara2008}%
  \BibitemOpen
  \bibfield  {author} {\bibinfo {author} {\bibnamefont {{Y. Kamihara, T.
  Watanabe, M. Hirano, and H. Hosono}}},\ }\href {\doibase 10.1021/ja800073m}
  {\bibfield  {journal} {\bibinfo  {journal} {J. Am. Chem. Soc.}\ }\textbf
  {\bibinfo {volume} {130}},\ \bibinfo {pages} {3296} (\bibinfo {year}
  {2008})}\BibitemShut {NoStop}%
\bibitem [{\citenamefont {Alloul}\ \emph {et~al.}(2009)\citenamefont {Alloul},
  \citenamefont {Bobroff}, \citenamefont {Gabay},\ and\ \citenamefont
  {Hirschfeld}}]{Alloul2009}%
  \BibitemOpen
  \bibfield  {author} {\bibinfo {author} {\bibfnamefont {H.}~\bibnamefont
  {Alloul}}, \bibinfo {author} {\bibfnamefont {J.}~\bibnamefont {Bobroff}},
  \bibinfo {author} {\bibfnamefont {M.}~\bibnamefont {Gabay}}, \ and\ \bibinfo
  {author} {\bibfnamefont {P.~J.}\ \bibnamefont {Hirschfeld}},\ }\href
  {\doibase 10.1103/RevModPhys.81.45} {\bibfield  {journal} {\bibinfo
  {journal} {Rev. Mod. Phys.}\ }\textbf {\bibinfo {volume} {81}},\ \bibinfo
  {pages} {45} (\bibinfo {year} {2009})}\BibitemShut {NoStop}%
\bibitem [{\citenamefont {Wadati}\ \emph {et~al.}(2010)\citenamefont {Wadati},
  \citenamefont {Elfimov},\ and\ \citenamefont {Sawatzky}}]{Wadati2010}%
  \BibitemOpen
  \bibfield  {author} {\bibinfo {author} {\bibfnamefont {H.}~\bibnamefont
  {Wadati}}, \bibinfo {author} {\bibfnamefont {I.}~\bibnamefont {Elfimov}}, \
  and\ \bibinfo {author} {\bibfnamefont {G.~A.}\ \bibnamefont {Sawatzky}},\
  }\href {\doibase 10.1103/PhysRevLett.105.157004} {\bibfield  {journal}
  {\bibinfo  {journal} {Phys. Rev. Lett.}\ }\textbf {\bibinfo {volume} {105}},\
  \bibinfo {pages} {157004} (\bibinfo {year} {2010})}\BibitemShut {NoStop}%
\bibitem [{\citenamefont {{A. S. Sefat, R. Jin, M. A. McGuire, B. C. Sales, D.
  J. Singh, and D. Mandrus}}(2008)}]{Sefat2008}%
  \BibitemOpen
  \bibfield  {author} {\bibinfo {author} {\bibnamefont {{A. S. Sefat, R. Jin,
  M. A. McGuire, B. C. Sales, D. J. Singh, and D. Mandrus}}},\ }\href {\doibase
  10.1103/PhysRevLett.101.117004} {\bibfield  {journal} {\bibinfo  {journal}
  {Phys. Rev. Lett.}\ }\textbf {\bibinfo {volume} {101}},\ \bibinfo {pages}
  {117004} (\bibinfo {year} {2008})}\BibitemShut {NoStop}%
\bibitem [{\citenamefont {{L. J. Li, Y. K. Luo, Q. B. Wang, H. Chen, Z. Ren, Q.
  Tao, Y. K. Li, X. Lin, M. He, Z. W. Zhu, G. H. Cao and Z. A.
  Xu}}(2009)}]{Li2009NiBaFe2As2}%
  \BibitemOpen
  \bibfield  {author} {\bibinfo {author} {\bibnamefont {{L. J. Li, Y. K. Luo,
  Q. B. Wang, H. Chen, Z. Ren, Q. Tao, Y. K. Li, X. Lin, M. He, Z. W. Zhu, G.
  H. Cao and Z. A. Xu}}},\ }\href
  {http://stacks.iop.org/1367-2630/11/i=2/a=025008} {\bibfield  {journal}
  {\bibinfo  {journal} {New J. Phys.}\ }\textbf {\bibinfo {volume} {11}},\
  \bibinfo {pages} {025008} (\bibinfo {year} {2009})}\BibitemShut {NoStop}%
\bibitem [{\citenamefont {{A. Thaler, , H. Hodovanets, M. S. Torikachvili, S.
  Ran, A. Kracher, W. Straszheim, J. Q. Yan, E. Mun, and P. C.
  Canfield}}(2011)}]{Thaler2011Mn}%
  \BibitemOpen
  \bibfield  {author} {\bibinfo {author} {\bibnamefont {{A. Thaler, , H.
  Hodovanets, M. S. Torikachvili, S. Ran, A. Kracher, W. Straszheim, J. Q. Yan,
  E. Mun, and P. C. Canfield}}},\ }\href {\doibase 10.1103/PhysRevB.84.144528}
  {\bibfield  {journal} {\bibinfo  {journal} {Phys. Rev. B}\ }\textbf {\bibinfo
  {volume} {84}},\ \bibinfo {pages} {144528} (\bibinfo {year}
  {2011})}\BibitemShut {NoStop}%
\bibitem [{\citenamefont {{J. E. Hoffman}}(2011)}]{Hoffman-review-2011}%
  \BibitemOpen
  \bibfield  {author} {\bibinfo {author} {\bibnamefont {{J. E. Hoffman}}},\
  }\href {\doibase 10.1088/0034-4885/74/12/124513} {\bibfield  {journal}
  {\bibinfo  {journal} {Rep. Prog. Phys.}\ }\textbf {\bibinfo {volume} {74}},\
  \bibinfo {pages} {124513} (\bibinfo {year} {2011})}\BibitemShut {NoStop}%
\bibitem [{\citenamefont {{J. H. Tapp, Z. Tang, B. Lv, K. Sasmal, B. Lorenz, P.
  C. W. Chu, and A. M. Guloy}}(2008)}]{Tapp2008}%
  \BibitemOpen
  \bibfield  {author} {\bibinfo {author} {\bibnamefont {{J. H. Tapp, Z. Tang,
  B. Lv, K. Sasmal, B. Lorenz, P. C. W. Chu, and A. M. Guloy}}},\ }\href
  {\doibase 10.1103/PhysRevB.78.060505} {\bibfield  {journal} {\bibinfo
  {journal} {Phys. Rev. B}\ }\textbf {\bibinfo {volume} {78}},\ \bibinfo
  {pages} {060505} (\bibinfo {year} {2008})}\BibitemShut {NoStop}%
\bibitem [{\citenamefont {{X.C. Wang and Q.Q. Liu and Y.X. Lv and W.B. Gao and
  L.X. Yang and R.C. Yu and F.Y. Li and C.Q. Jin}}(2008)}]{Wang2008}%
  \BibitemOpen
  \bibfield  {author} {\bibinfo {author} {\bibnamefont {{X.C. Wang and Q.Q. Liu
  and Y.X. Lv and W.B. Gao and L.X. Yang and R.C. Yu and F.Y. Li and C.Q.
  Jin}}},\ }\href {\doibase 10.1016/j.ssc.2008.09.057} {\bibfield  {journal}
  {\bibinfo  {journal} {Solid State Commun.}\ }\textbf {\bibinfo {volume}
  {148}},\ \bibinfo {pages} {538 } (\bibinfo {year} {2008})}\BibitemShut
  {NoStop}%
\bibitem [{\citenamefont {{M. J. Pitcher , D. R. Parker , P. Adamson, S. J. C.
  Herkelrath, A. T. Boothroyd, R. M. Ibberson , M. Brunelli, and S. J.
  Clarke}}(2008)}]{Pitcher2008}%
  \BibitemOpen
  \bibfield  {author} {\bibinfo {author} {\bibnamefont {{M. J. Pitcher , D. R.
  Parker , P. Adamson, S. J. C. Herkelrath, A. T. Boothroyd, R. M. Ibberson ,
  M. Brunelli, and S. J. Clarke}}},\ }\href {\doibase 10.1039/B813153H}
  {\bibfield  {journal} {\bibinfo  {journal} {Chem. Commun.}\ ,\ \bibinfo
  {pages} {5918}} (\bibinfo {year} {2008})}\BibitemShut {NoStop}%
\bibitem [{\citenamefont {{F.C. Hsu, J.-Y. Luo, K.-W. Yeh, T.-K. Chen, T.-W.
  Huang, P. M. Wu, Y.-C. Lee, Y.-L. Huang, Y.-Y. Chu, D.-C. Yan, and M.-K.
  Wu}}(2008)}]{Hsu2008}%
  \BibitemOpen
  \bibfield  {author} {\bibinfo {author} {\bibnamefont {{F.C. Hsu, J.-Y. Luo,
  K.-W. Yeh, T.-K. Chen, T.-W. Huang, P. M. Wu, Y.-C. Lee, Y.-L. Huang, Y.-Y.
  Chu, D.-C. Yan, and M.-K. Wu}}},\ }\href {\doibase 10.1073/pnas.0807325105}
  {\bibfield  {journal} {\bibinfo  {journal} {Proc. Natl. Acad. Sci.}\ }\textbf
  {\bibinfo {volume} {105}},\ \bibinfo {pages} {14262} (\bibinfo {year}
  {2008})}\BibitemShut {NoStop}%
\bibitem [{\citenamefont {{W. Li, H. Ding, P. Deng, K. Chang, C. Song, K. He,
  L. Wang, X. Ma, J.-P. Hu, X. Chen, and Q.-K. Xue}}(2011)}]{Li2011}%
  \BibitemOpen
  \bibfield  {author} {\bibinfo {author} {\bibnamefont {{W. Li, H. Ding, P.
  Deng, K. Chang, C. Song, K. He, L. Wang, X. Ma, J.-P. Hu, X. Chen, and Q.-K.
  Xue}}},\ }\href {\doibase 10.1038/nphys2155} {\bibfield  {journal} {\bibinfo
  {journal} {Nature Physics}\ }\textbf {\bibinfo {volume} {8}},\ \bibinfo
  {pages} {126} (\bibinfo {year} {2011})}\BibitemShut {NoStop}%
\bibitem [{\citenamefont {{S. Chi, S. Grothe, R. Liang, P. Dosanjh, W. N.
  Hardy, S. A. Burke, D. A. Bonn, Y. Pennec}}(2012)}]{Chi2012}%
  \BibitemOpen
  \bibfield  {author} {\bibinfo {author} {\bibnamefont {{S. Chi, S. Grothe, R.
  Liang, P. Dosanjh, W. N. Hardy, S. A. Burke, D. A. Bonn, Y. Pennec}}},\
  }\href {\doibase 10.1103/PhysRevLett.109.087002} {\bibfield  {journal}
  {\bibinfo  {journal} {Phys. Rev. Lett.}\ }\textbf {\bibinfo {volume} {109}},\
  \bibinfo {pages} {087002} (\bibinfo {year} {2012})}\BibitemShut {NoStop}%
\bibitem [{\citenamefont {{T. Hanaguri, K. Kitagawa, K. Matsubayashi, Y.
  Mazaki, Y. Uwatoko, and H. Takagi}}(2012)}]{Hanaguri2012}%
  \BibitemOpen
  \bibfield  {author} {\bibinfo {author} {\bibnamefont {{T. Hanaguri, K.
  Kitagawa, K. Matsubayashi, Y. Mazaki, Y. Uwatoko, and H. Takagi}}},\ }\href
  {\doibase 10.1103/PhysRevB.85.214505} {\bibfield  {journal} {\bibinfo
  {journal} {Phys. Rev. B}\ }\textbf {\bibinfo {volume} {85}},\ \bibinfo
  {pages} {214505} (\bibinfo {year} {2012})}\BibitemShut {NoStop}%
\bibitem [{\citenamefont {{M. P. Allan, A. W. Rost, A. P. Mackenzie, Yang Xie,
  J. C. Davis, K. Kihou, C. H. Lee, A. Iyo, H. Eisaki, T.-M.
  Chuang}}(2012)}]{Allan2012}%
  \BibitemOpen
  \bibfield  {author} {\bibinfo {author} {\bibnamefont {{M. P. Allan, A. W.
  Rost, A. P. Mackenzie, Yang Xie, J. C. Davis, K. Kihou, C. H. Lee, A. Iyo, H.
  Eisaki, T.-M. Chuang}}},\ }\href {\doibase 10.1126/science.1218726}
  {\bibfield  {journal} {\bibinfo  {journal} {Science}\ }\textbf {\bibinfo
  {volume} {336}},\ \bibinfo {pages} {563} (\bibinfo {year}
  {2012})}\BibitemShut {NoStop}%
\bibitem [{\citenamefont {{T. H\"anke, S. Sykora, R. Schlegel, D. Baumann, L.
  Harnagea, S. Wurmehl, M. Daghofer, B. B\"uchner, J. van den Brink, and C.
  Hess}}(2012)}]{Hess-STM}%
  \BibitemOpen
  \bibfield  {author} {\bibinfo {author} {\bibnamefont {{T. H\"anke, S. Sykora,
  R. Schlegel, D. Baumann, L. Harnagea, S. Wurmehl, M. Daghofer, B. B\"uchner,
  J. van den Brink, and C. Hess}}},\ }\href {\doibase
  10.1103/PhysRevLett.108.127001} {\bibfield  {journal} {\bibinfo  {journal}
  {Phys. Rev. Lett.}\ }\textbf {\bibinfo {volume} {108}},\ \bibinfo {pages}
  {127001} (\bibinfo {year} {2012})}\BibitemShut {NoStop}%
\bibitem [{\citenamefont {{C. L. Song, Y.-L.Wang, P. Cheng, Y.-P. Jiang, W. Li,
  T. Zhang, Z. Li, K. He, L. Wang, J.-F. Jia, H.-H. Hung, C. Wu, X. Ma, X.
  Chen, and Q.-K. Xu}}(2011)}]{Song2011}%
  \BibitemOpen
  \bibfield  {author} {\bibinfo {author} {\bibnamefont {{C. L. Song, Y.-L.Wang,
  P. Cheng, Y.-P. Jiang, W. Li, T. Zhang, Z. Li, K. He, L. Wang, J.-F. Jia,
  H.-H. Hung, C. Wu, X. Ma, X. Chen, and Q.-K. Xu}}},\ }\href {\doibase
  10.1126/science.1202226} {\bibfield  {journal} {\bibinfo  {journal}
  {Science}\ }\textbf {\bibinfo {volume} {332}},\ \bibinfo {pages} {1410}
  (\bibinfo {year} {2011})}\BibitemShut {NoStop}%
\bibitem [{\citenamefont {Berciu}\ \emph {et~al.}(2009)\citenamefont {Berciu},
  \citenamefont {Elfimov},\ and\ \citenamefont {Sawatzky}}]{Berciu2009}%
  \BibitemOpen
  \bibfield  {author} {\bibinfo {author} {\bibfnamefont {M.}~\bibnamefont
  {Berciu}}, \bibinfo {author} {\bibfnamefont {I.}~\bibnamefont {Elfimov}}, \
  and\ \bibinfo {author} {\bibfnamefont {G.~A.}\ \bibnamefont {Sawatzky}},\
  }\href {\doibase 10.1103/PhysRevB.79.214507} {\bibfield  {journal} {\bibinfo
  {journal} {Phys. Rev. B}\ }\textbf {\bibinfo {volume} {79}},\ \bibinfo
  {pages} {214507} (\bibinfo {year} {2009})}\BibitemShut {NoStop}%
\bibitem [{\citenamefont {{S. V. Borisenko, V. B. Zabolotnyy, A. A. Kordyuk, D.
  V. Evtushinsky, T. K. Kim, I. V. Morozov, R. Follath and B.
  B\"uchner}}(2012)}]{Borisenko2012}%
  \BibitemOpen
  \bibfield  {author} {\bibinfo {author} {\bibnamefont {{S. V. Borisenko, V. B.
  Zabolotnyy, A. A. Kordyuk, D. V. Evtushinsky, T. K. Kim, I. V. Morozov, R.
  Follath and B. B\"uchner}}},\ }\href {\doibase 10.3390/sym4010251} {\bibfield
   {journal} {\bibinfo  {journal} {Symmetry}\ }\textbf {\bibinfo {volume}
  {4}},\ \bibinfo {pages} {251} (\bibinfo {year} {2012})}\BibitemShut {NoStop}%
\bibitem [{\citenamefont {{J. Ferber, K. Foyevtsova, R. Valent{\'\i} and H. O.
  Jeschke}}(2012)}]{Ferber2012}%
  \BibitemOpen
  \bibfield  {author} {\bibinfo {author} {\bibnamefont {{J. Ferber, K.
  Foyevtsova, R. Valent{\'\i} and H. O. Jeschke}}},\ }\href {\doibase
  10.1103/PhysRevB.85.094505} {\bibfield  {journal} {\bibinfo  {journal} {Phys.
  Rev. B}\ }\textbf {\bibinfo {volume} {85}},\ \bibinfo {pages} {094505}
  (\bibinfo {year} {2012})}\BibitemShut {NoStop}%
\bibitem [{\citenamefont {{T. Hajiri, T. Ito, R. Niwa, M. Matsunami, B. H. Min,
  Y. S. Kwon, and S. Kimura}}(2012)}]{Hajiri2012}%
  \BibitemOpen
  \bibfield  {author} {\bibinfo {author} {\bibnamefont {{T. Hajiri, T. Ito, R.
  Niwa, M. Matsunami, B. H. Min, Y. S. Kwon, and S. Kimura}}},\ }\href
  {\doibase 10.1103/PhysRevB.85.094509} {\bibfield  {journal} {\bibinfo
  {journal} {Phys. Rev. B}\ }\textbf {\bibinfo {volume} {85}},\ \bibinfo
  {pages} {094509} (\bibinfo {year} {2012})}\BibitemShut {NoStop}%
\bibitem [{\citenamefont {{J. Tersoff and D. R. Hamann}}(1983)}]{Tersoff1998}%
  \BibitemOpen
  \bibfield  {author} {\bibinfo {author} {\bibnamefont {{J. Tersoff and D. R.
  Hamann}}},\ }\href {\doibase 10.1103/PhysRevLett.50.1998} {\bibfield
  {journal} {\bibinfo  {journal} {Phys. Rev. Lett.}\ }\textbf {\bibinfo
  {volume} {50}},\ \bibinfo {pages} {1998} (\bibinfo {year}
  {1983})}\BibitemShut {NoStop}%
\bibitem [{Note1()}]{Note1}%
  \BibitemOpen
  \bibinfo {note} {The energetic positions of bound-state resonances are not
  expected to be influenced by the tip state, which we have confirmed by
  comparing the data shown here for the Fe-D$_2$-1 with 4 K spectra on this
  defect from our previous data \cite {Chi2012} showing the lower weight of
  $\Delta _2$.}\BibitemShut {Stop}%
\bibitem [{Note2()}]{Note2}%
  \BibitemOpen
  \bibinfo {note} {We note that this normalization procedure may slightly shift
  energetic features especially if there is an overall redistribution of
  spectral weight, a gap suppression, or multiple peaks that are within our 200
  $\mu $V resolution. Since these factors likely exceed the small statistical
  uncertainty of locating the peak position, we report only approximate values
  for bound state energies.}\BibitemShut {Stop}%
\bibitem [{\citenamefont {{H. Amara, S. Latil, V. Meunier, Ph. Lambin, and
  J.-C. Charlier}}(2007)}]{Amara2007}%
  \BibitemOpen
  \bibfield  {author} {\bibinfo {author} {\bibnamefont {{H. Amara, S. Latil, V.
  Meunier, Ph. Lambin, and J.-C. Charlier}}},\ }\href {\doibase
  10.1103/PhysRevB.76.115423} {\bibfield  {journal} {\bibinfo  {journal} {Phys.
  Rev. B}\ }\textbf {\bibinfo {volume} {76}},\ \bibinfo {pages} {115423}
  (\bibinfo {year} {2007})}\BibitemShut {NoStop}%
\bibitem [{\citenamefont {{J. Ishioka, T. Fujii, K. Katono, K. Ichimura, T.
  Kurosawa, M. Oda, and S. Tanda}}(2011)}]{Ishioka2011}%
  \BibitemOpen
  \bibfield  {author} {\bibinfo {author} {\bibnamefont {{J. Ishioka, T. Fujii,
  K. Katono, K. Ichimura, T. Kurosawa, M. Oda, and S. Tanda}}},\ }\href
  {\doibase 10.1103/PhysRevB.84.245125} {\bibfield  {journal} {\bibinfo
  {journal} {Phys. Rev. B}\ }\textbf {\bibinfo {volume} {84}},\ \bibinfo
  {pages} {245125} (\bibinfo {year} {2011})}\BibitemShut {NoStop}%
\bibitem [{\citenamefont {{H. Huang, Y. Gao, D. Zhang, and C. S.
  Ting}}(2011)}]{Huang2011}%
  \BibitemOpen
  \bibfield  {author} {\bibinfo {author} {\bibnamefont {{H. Huang, Y. Gao, D.
  Zhang, and C. S. Ting}}},\ }\href {\doibase 10.1103/PhysRevB.84.134507}
  {\bibfield  {journal} {\bibinfo  {journal} {Phys. Rev. B}\ }\textbf {\bibinfo
  {volume} {84}},\ \bibinfo {pages} {134507} (\bibinfo {year}
  {2011})}\BibitemShut {NoStop}%
\bibitem [{\citenamefont {{Y. Inoue, Y. Yamakawa, and H.
  Kontani}}(2012)}]{Inoue2012}%
  \BibitemOpen
  \bibfield  {author} {\bibinfo {author} {\bibnamefont {{Y. Inoue, Y. Yamakawa,
  and H. Kontani}}},\ }\href {\doibase 10.1103/PhysRevB.85.224506} {\bibfield
  {journal} {\bibinfo  {journal} {Phys. Rev. B}\ }\textbf {\bibinfo {volume}
  {85}},\ \bibinfo {pages} {224506} (\bibinfo {year} {2012})}\BibitemShut
  {NoStop}%
\bibitem [{\citenamefont {{A. V. Balatsky, I. Vekhter and J.-X.
  Zhu}}(2006)}]{Balatsky2006review}%
  \BibitemOpen
  \bibfield  {author} {\bibinfo {author} {\bibnamefont {{A. V. Balatsky, I.
  Vekhter and J.-X. Zhu}}},\ }\href {\doibase 10.1103/RevModPhys.78.373}
  {\bibfield  {journal} {\bibinfo  {journal} {Rev. Mod. Phys.}\ }\textbf
  {\bibinfo {volume} {78}},\ \bibinfo {pages} {373} (\bibinfo {year}
  {2006})}\BibitemShut {NoStop}%
\end{thebibliography}%

\end{document}